\documentclass[fleqn,usenatbib]{mnras}
\usepackage{newtxtext,newtxmath}
\usepackage[T1]{fontenc}
 \usepackage{subcaption}

\usepackage{amsmath}
\usepackage{graphicx}
\usepackage{amsmath}
\usepackage{xspace} 
\newcommand{\teff}{\ensuremath{T_{\mathrm{eff}}}\xspace}

\newcommand{\logg}{\ensuremath{\log g}\xspace}
\newcommand{\feh}{\rm{[Fe/H]}\xspace}
\newcommand{\cfe}{\rm{[C/Fe]}\xspace}

\newcommand{\alphafe}{\rm{[\ensuremath{\alpha}/Fe]}\xspace}

\title[Information content of BP/RP spectra]{Information content of BP/RP spectra in Gaia DR3}

\author[C. E. C. Witten et al.]{Callum E. C. Witten,$^{1}$\thanks{E-mail: cw795@cam.ac.uk}
David S. Aguado,$^{2,3}$
Jason L. Sanders,$^{4}$
Vasily Belokurov,$^{1,5}$
N. Wyn Evans,$^{1}$
\newauthor{Sergey E. Koposov,$^{6,1}$
Carlos Allende Prieto,$^{7,8}$
Francesca De Angeli $^{1}$ and
Mike J. Irwin$^{1}$}
\\
$^{1}$Institute of Astronomy, University of Cambridge, Madingley Road, Cambridge CB3 0HA, United Kingdom\\
$^{2}$Dipartimento di Fisica e Astronomia, Universit \'a degli Studi di Firenze, Via G. Sansone 1, I-50019 Sesto Fiorentino, Italy \\
$^{3}$INAF/Osservatorio Astrofisico di Arcetri, Largo E. Fermi 5, I-50125 Firenze, Italy\\
$^{4}$University College London, Gower St., London WC1E 6BT, UK\\
$^5$Center for Computational Astrophysics, Flatiron Institute, 162 5th Avenue, New York, NY 10010, USA\\
$^{6}$Institute for Astronomy, University of Edinburgh, Royal Observatory, Blackford Hill, Edinburgh EH9 3HJ, UK\\
$^{7}$Instituto de Astrofísica de Canarias, Vía Láctea, 38205 La Laguna, Tenerife, Spain\\
$^{8}$Universidad de La Laguna, Departamento de Astrofísica, 38206 La Laguna, Tenerife, Spain
}

\date{Accepted XXX. Received YYY; in original form ZZZ}

\pubyear{2022}

\begin{document}
\label{firstpage}
\pagerange{\pageref{firstpage}--\pageref{lastpage}}
\maketitle

\begin{abstract}
Gaia Data Release 3 has provided the astronomical community with the largest stellar spectroscopic survey to date ($>$ 220 million sources). The low resolution (R$\sim$50) blue photometer (BP) and red photometer (RP) spectra will allow for the estimation of stellar atmospheric parameters such as effective temperature, surface gravity and metallicity. We create mock Gaia BP/RP spectra and use Fisher information matrices to probe the resolution limit of stellar parameter measurements using BP/RP spectra. The best-case scenario uncertainties that this analysis provides are then used to produce a mock-observed stellar population in order to evaluate the false positive rate (FPR) of identifying extremely metal-poor (EMP) stars. We conclude that the community will be able to confidently identify metal-poor stars at magnitudes brighter than $G = 16$ using BP/RP spectra. At fainter magnitudes true detections will start to be overwhelmed by false positives. When adopting the commonly-used $G < 14$ limit for metal-poor star searches, we find a FPR for the low-metallicity regimes [Fe/H] $<$ -2, -2.5 and -3 of just 14$\%$, 33$\%$ and 56$\%$ respectively, offering the potential for significant improvements on previous targeting campaigns. Additionally, we explore the chemical sensitivity obtainable directly from BP/RP spectra for Carbon and $\alpha$-elements. We find an absolute Carbon abundance uncertainty of $\sigma_{A(C)} < 1$ dex for Carbon-enriched metal-poor (CEMP) stars, indicating the potential to identify a CEMP stellar population for follow-up confirmation with higher resolution spectroscopy. Finally, we find that large uncertainties in $\alpha$-element abundance measurements using BP/RP spectra means that efficiently obtaining these abundances will be challenging.
\end{abstract}


\section{Introduction}

The {\it Gaia} satellite was launched by the European Space Agency in 2013 and is expected to continue data-taking until 2026~\citep{Pr16}. The main objective of {\it Gaia} is micro-arcsecond astrometry, but it also collects stellar spectra using the red photometer (RP), blue photometer (BP), and radial velocity spectrometer (RVS). The RVS spectra are only publicly available for stars brighter than $G_{\rm RVS}=12$, which is about 7.2 million stars in {\it Gaia} Data Release 2~\citep{Ka19}.

However, on 13 June 2022, Data Release 3 (DR3) will provide for the first time low-resolution BP and RP spectra for hundreds of millions of stars. The combined BP/RP spectra cover the wavelength range 3300-10500 Å with a resolution between 13 and 85~\citep{Carrasco21}. These spectra have resolutions that are too poor to allow us to measure individual spectral lines. Nonetheless, the combined BP/RP spectra may be used to estimate some stellar properties such as effective temperature $T_{\rm eff}$, surface gravity $\log g$, and metallicity [M/H], and hence extract individual objects for further study. Given the huge size of the BP/RP dataset, this will likely contain representatives of extreme objects. We are therefore interested in devising algorithms to extract such objects.

Low metallicity stars in the Milky Way have been the subject of many recent searches~\citep{Venn04,Beers2005,Frebel15}. They are primitive objects, probably the descendants of Population III stars -- the very first, almost metal-free, objects that seeded chemical evolution in the Universe. \citet{Beers2005} introduced the nomenclature of extremely metal poor (EMP) stars to describe those stars with iron abundances below $1/1000$ of the Solar, or [Fe/H]$<$-3. Targeting strategies have been devised to hunt down bright candidates from public surveys like 2MASS near-infrared and WISE mid-infrared photometry for subsequent spectroscopic follow-up~\citep[e.g.,][]{Sc14,Li21}. Dedicated surveys using narrow band imaging based around the metallicity sensitive near-infrared Ca H{\&}K lines are underway~\cite[the PRISTINE survey,][]{St17} with the goal of uncovering hundreds of candidates~\citep[e.g.,][]{St18,Ve20}. At such low metallicities, the fraction of the carbon-enhanced metal-poor (CEMP) stars dramatically increases~\citep{Ca12,Sa15,Yoon2016}. Metal-poor CEMP stars have [C/Fe] > +1, so that compared to the Sun they are carbon enhanced at least ten times more than iron. CEMP stars are often further subdivided as to whether r-process or s-process elements are enhanced, while CEMP-no stars have no enhancement.

An earlier study \citep{allende16} hinted that Gaia's low-dispersion spectra are quite useful to constrain atmospheric parameters down to very low metallicity\footnote{\citet{allende16} indicates that photometric colors can be used, in some instances, to constrain stellar parameters, one could therefore use the publicly avalaible GaiaXPy code (\url{https://gaia-dpci.github.io/GaiaXPy-website/}) to simulate Gaia photometry in order to confirm this, however, it cannot yet be used to simulate Gaia BP/RP spectra.}. The aim of this paper is therefore to examine whether EMP and CEMP stars can be extracted in an efficient manner from the BP/RP spectral database. The paper is arranged as follows. Section~\ref{sec:Mock spectra} describes our methodology for creation of mock BP/RP spectra for Galactic populations. Then, in Section~\ref{sec:Analysis} we exploit the Fisher matrix to understand the degeneracies between extracted parameters in these low resolution spectra. Finally, we outline the resolution of Gaia BP/RP in stellar parameter measurements and additionally its ability to efficiently extract EMP candidates.

\section{Mock Gaia BP/RP spectra}

\label{sec:Mock spectra}

In order to probe the potential resolution of stellar parameter measurements using Gaia spectra, we must first produce a mock Gaia BP/RP spectra catalogue. The form of these spectra are dictated by the instrumentation of the Gaia spectrophotometers, such as their transmission curves, spectral resolution, original sampling, and observed line spread function (LSF). According to these parameters we built simulated Gaia BP/RP spectra from existing spectral libraries from the literature. 

\subsection{Stellar models}
The selected stellar models came from two different sources, the PHOENIX \citep{husser13} and \citet{aguado17b} libraries. These two sets of synthetic stellar spectra cover different aspects needed for this study. The PHOENIX library uses 1D radiative hydrodynamic models (spherically-symetric), leading to a more realistic modelling of stellar atmospheres in low-gravity giant populations, similar to the well known MARCS library \citep{gustafsson08}\footnote{It is important to note the MARCS library does not extend the spherically-symetric models to dwarf populations while our selected PHOENIX library actually does.}. However, our choice is further supported by the fact the PHOENIX library covers a wide range in $\alpha$-elements abundance. On the other hand, the library by \citet{aguado17b} using plane-parallel symmetry (1D) will allow us to include synthetic models for carbon-enhanced metal-poor (CEMP) stars. This set of stellar models share the main properties of a broader synthetic library from \citet{AllendePrieto18}.
 
 \subsubsection{The PHOENIX library}\label{sec:PHX}
 This library, made publicly available by \citet{husser13}, employs a grid of 1D stellar atmosphere models computed by PHOENIX code. The fact that these models use spherical symmetry make them very suitable for evolved phases of stellar evolution, especially for cool giants. The synthesis is also performed by a particular mode available in the PHOENIX code. Every model is synthesised by assuming a unique microturbulence derived from convection velocities \citep{ludwig1999} within the atmospheric model and it is directly related with the large scale motion in the stellar atmosphere (macroturbulence). We consider this a priori approach an advantage of the PHOENIX library. Remarkably, the authors included some line profiles for neutral atoms of lighter elements in a non thermodynamical equilibrium (NLTE). A more compact version of the library, is available from the PHOENIX webpage\footnote{\url{http://phoenix.astro.physik.uni-goettingen.de/}},  smoothed to the X-SHOOTER resolution and covering the following parameters range:
 \begin{itemize}
\item $2300~\mathrm{K} < \teff < 8000~\mathrm{K}$, $\Delta \teff = 100~\mathrm{K}$
\item $-0.5  < \logg < 7.0$, $\Delta \logg = 0.5$
\item $-4.0  < \feh < +1.0$, $\Delta \feh = 0.5$
\item $-0.4  < \alphafe < +1.0$, $\Delta \alphafe = 0.2$
\end{itemize}
We download the data and build a super 4D cube of models in a suitable manner for our purposes (see Sec. \ref{sec:inter})  
 
    \subsubsection{The \citet{aguado17b} library}\label{sec:Carbon Grid}
For this work we use a grid of synthetic models that is an extension in metallicity and temperature to those publicly available from \citet{aguado17b}. This library is based on a grid of \citet{kur79} atmospheric models computed with ATLAS12 \citep{sbordone07} assuming a 1D-LTE approach. The synthesis is performed with the radiative transfer code ASSET \citep{koe08}.  These spectra have an assumed $[\alpha /\rm Fe] = 0.4$ - a typical value for metal-poor halo stars \citep[see e.g.,][]{Sneden1991, Tomkin1992} and microturbulance of $\xi=2.0$\,km/s. This library covers a wider range of temperatures than those available in the PHOENIX or MARCS collections and focuses on FGK-type. Additionally, it includes very recent and accurate continuum opacities.
Although for the parametric study we present in this work we mostly focus on the PHOENIX library, the chance to extend this library and incorporate carbon abundance in the synthesis of the spectra and also in their models of stellar atmospheres makes the the \citet{aguado17b} library of great use for this analysis. The parameters covered by the grid are summarised as follows:

 \begin{itemize}
\item $4500~\mathrm{K} < \teff < 7000~\mathrm{K}$, $\Delta \teff = 250~\mathrm{K}$
\item $1.0  < \logg < 5.0$, $\Delta \logg = 0.5$
\item $-4.0  < \feh < +1.0$, $\Delta \feh = 0.5$
\item $-1.0  < \cfe < +3.0$, $\Delta \cfe = 1.0$
\end{itemize}

This carbon-grid was designed to account for metal-poor stars with carbon enrichment and has been tested in recent works \citep[see, e.g.][]{Aguado2021a, Aguado2021b}.
 
\subsubsection{The FERRE interpolation}\label{sec:inter}

To prepare both libraries to be converted to Gaia BP/RP shape we firstly interpolate to produce a PHOENIX grid with finer steps in both \logg and \feh  of 0.25 dex and 0.1 dex respectively. Accordingly, we also interpolated the carbon-grid in \feh and \cfe to steps of 0.1 and 0.25 dex respectively. This interpolation produces no additional information but allows us to produce smoother variations across changing stellar parameters. This interpolation was completed using FERRE\footnote{{\tt FERRE} is available from \url{http://github.com/callendeprieto/ferre}} from \citet{alle06}. We use a cubic B$\rm\Acute{e}$zier interpolation, where FERRE operates over the nodes of the closest stellar parameter values and is able to interpolate simultaneously across the entire parameter space. This arrives at a global solution that is considerably smoother than a linear interpolation. 
We then restricted the wavelength coverage to between 300-1100\,nm and smoothed the models to a resolving power of $\rm R=1000$.
 
\subsection{BP/RP simulation}
\label{sec:BP RP Simulation}
A large set of calibrators were used to produce passbands released alongside the Early Data Release 3 (EDR3) \citep[see][]{Riello21}. We apply these photon transmission curves, $T_\mathrm{phot}(\lambda)$, to our raw synthetic spectra, $f(\lambda)\:($erg$\,$cm$^{-2}\,$s$^{-1}\,$\AA$^{-1})$, to replicate the transmitted BP/RP flux in each pixel, $f_\mathrm{trans}($photons$\, $s$^{-1})$, 
\begin{equation}
    f_\mathrm{trans} = \int T_\mathrm{phot}(\lambda)\, \lambda\, f(\lambda)\, \mathrm{d}\lambda,
	\label{eq:transmitted flux}
\end{equation}
where the integration range is defined by the resolution element of the spectrograph. This resolution element is both a function of wavelength and the object's position in the Gaia focal plane \citep{Carrasco21}. The variation with position in the focal plane is however minimal and hence we fit a single curve to the data from \citet{Carrasco21} and use this to estimate the width of the wavelength bins. We start at the lowest wavelengths for which BP and RP have non-zero photon transmission, 325 nm and 600 nm respectively, taking the corresponding bin width for that minimum wavelength, building the following bins from the previous bin's upper bound wavelength. 

These steps replicate the bandwidths and wavelength resolutions of the Gaia BP and RP spectra. However we must also take into account the smearing effects of the LSF. The result of the LSF is that light of a given wavelength is registered in not just one wavelength bin/pixel but instead also in the neighbouring pixels. In applying this LSF to our transmitted spectrum, $f_\mathrm{trans}$, we create the final mock BP/RP spectrum, $f_{\lambda}$. 

\begin{figure*}
	\includegraphics[width=\textwidth]{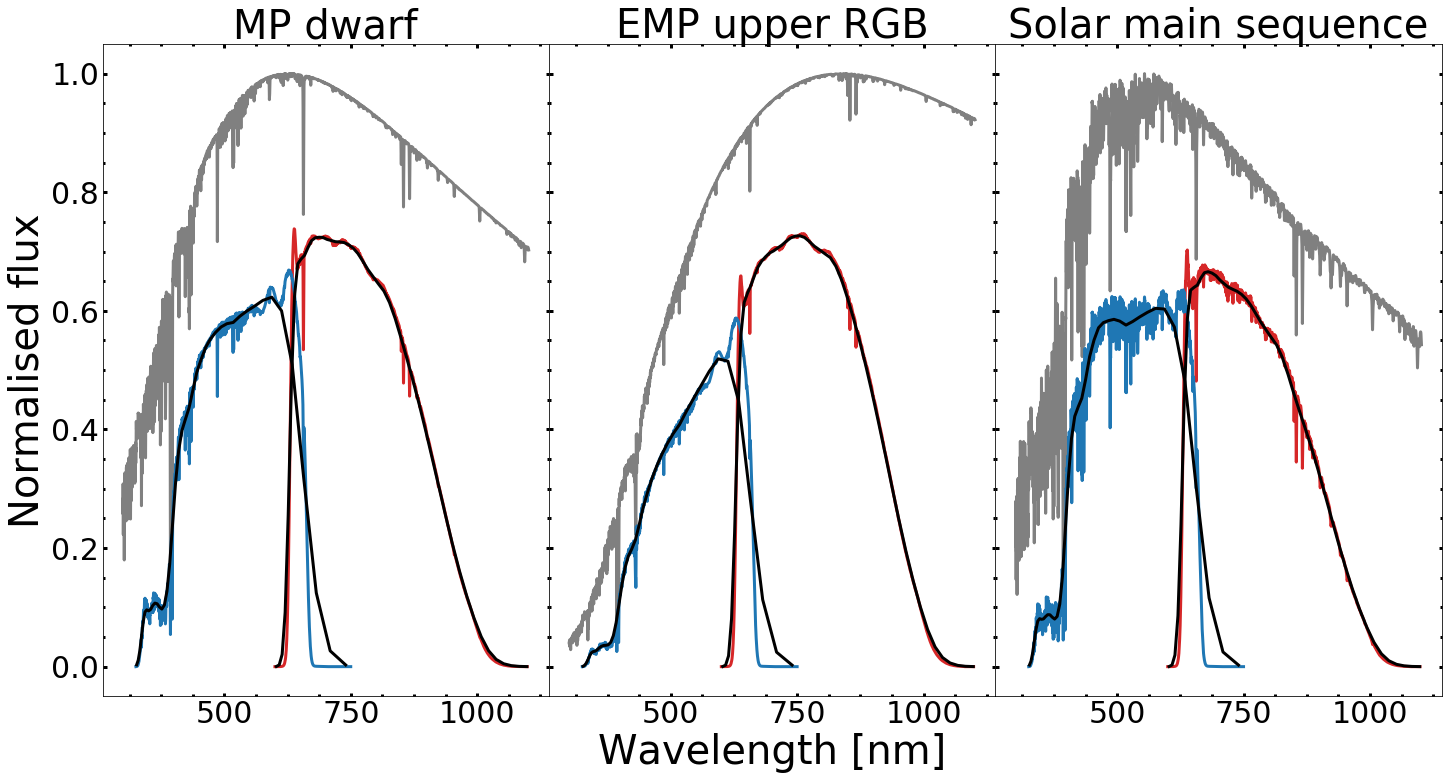}
    \caption{ The normalised photon count (counts $\rm s^{-1}$ $\rm nm^{-1}$) as a function of wavelength, showing the different steps taken in the production of the mock BP/RP spectra. The grey line shows the initial model spectra from \citet{AllendePrieto18}. The blue and red lines reveal the effect of applying the respective transmission curves of the BP and RP spectra. The black line shows the final mock observed BP/RP spectra after also reducing the resolution and applying the LSF. These are for three model stars, (left) a metal-poor dwarf ($T_\mathrm{eff}$ = 5500 K, $\log g$ = 4.4, [Fe/H] = -2), (middle) an extremely-metal-poor red-giant branch star ($T_\mathrm{eff}$ = 4500 K, $\log g$ = 1, [Fe/H] = -3) and (right) a typical Solar main sequence star ($T_\mathrm{eff}$ = 6000 K, $\log g$ = 4.5, [Fe/H] = 0). These spectra are in units of counts $\rm s^{-1}$ $\rm nm^{-1}$ for consistency, however for the final analysis of the BP/RP spectra we work in units of counts $\rm s^{-1}$.}
    \label{fig:SynthVsObs}
\end{figure*}

\citet{Carrasco21} quantifies the full-width half-maximum (FWHM) of the LSF to be $\sim 1$ BP/RP pixel, but this once again varies as a function of wavelength and focal plane position as well as the field of view and elapsed time throughout the mission. These many parameters are not possible to incorporate within our model and hence we make a tentative assumption, based on figures in \citet{Carrasco21}, to model the line spread function with a Gaussian with FWHM of 1.1 BP/RP pixels. We then convolve our transmitted flux by the LSF in order to produce a complete mock observed Gaia DR3 BP/RP stellar spectra. An example of the effect of each step in the process of creating the mock BP/RP spectra can be seen in Fig.~\ref{fig:SynthVsObs}. The final step taken to produce the BP/RP spectra is to normalize the flux by dividing through by the total $G$-band flux in the spectrum to ensure any change in total flux in the raw spectrum does not artificially create Fisher information. 

We note that the employed grids of synthetic spectra do not model for dust reddening. This is something we neglect and therefore recommend anyone using our analysis to dictate their observing strategy to focus on stars least affected by reddening. We have also considered the effects of a component of radial velocity with respect to Gaia on the observed BP/RP spectra. This could potentially act to shift spectroscopic structures into different wavelength bins than those we may ordinarily expect for a given set of stellar parameters. We ultimately find that for a star of high radial velocity ($V_{Rad} = 500$ km/s) we see no significant change in its spectra.

\section{Analysis of BP/RP spectra}
\label{sec:Analysis}

\subsection{Fisher information matrix}
\label{sec:Fisher}
Fisher information is the expectation value of the Hessian of the likelihood and has been shown to be useful in gauging the amount of information that an observable variable carries about the model parameters \citep[see][]{Fisher1935}. Fisher information matrix is computed at the point of maximum likelihood and is the inverse of the estimate of the covariance matrix \citep[see e.g.][]{Tegmark1997}. In our case the observable is the BP/RP spectrum and the parameters of interest are the numbers controlling the properties of the stellar atmosphere ($T_{\rm eff}$, log($g$) and [Fe/H]). We define our position in the stellar parameter space as the vector $\vec{x} = [T_{\rm{eff}},\rm{log}(g),\rm{[Fe/H]}]$.
We can use Fisher information matrices to indicate the information available at the position $\vec{x}$ as a function of wavelength, or alternatively, by summing the Fisher information matrices across the wavelength range we can obtain the total Fisher information for the spectrum. For example, assuming that the likelihood is Gaussian and unimodal, following \cite{Bonaca18}, the total Fisher information matrix at position $\vec{x}$ is defined as:

\begin{equation}
    \begin{aligned}
    \mathbf{F}(\vec{x}) &=\mathbf{C}(\vec{x})^{-1} = \left[\left(\frac{\partial \vec{f}}{\partial \vec{x}}\right)^\mathrm{T} \cdot \mathbf{C}_{f}^{-1} \cdot \left(\frac{\partial \vec{f}}{\partial \vec{x}}\right)\right] + \mathbf{V}^{-1} 
    \\ &= \sum_{\lambda} \left[\mathbf{F}(\vec{x},\lambda)\right] + \mathbf{V}^{-1} = \sum_{\lambda} \left[\left(\frac{\partial \vec{f}_{\lambda}}{\partial \vec{x}}\right)^\mathrm{T} \mathbf{C}_{f_{\lambda}}^{-1}\left(\frac{\partial \vec{f}_{\lambda}}{\partial \vec{x}}\right)\right] + \mathbf{V}^{-1},
    \end{aligned}
    \label{eq:Fisher}
\end{equation}

\noindent where $\mathbf{C}$ is the covariance matrix of the parameters we are interested in ($T_{\rm eff}$, log($g$) and [Fe/H]) at a fixed position in the stellar parameter space, $\vec{x}$. $\vec{f}$ denotes the vector flux across the wavelength space, $\mathbf{C}_{f}$ is the covariance matrix of the flux across the wavelength space, and finally, $\mathbf{V}$ is the covariance matrix of any priors on the measurement of the stellar parameters $T_{\rm{eff}},\rm{log}(g)$ and $\rm{[Fe/H]}$. In the second line of equation~\eqref{eq:Fisher} we express the dot product in summation notation, indicating how we can extract the Fisher information at a given wavelength, $\mathbf{F}(\vec{x},\lambda)$.

We calculate the differential of flux with respect to the position in the stellar parameter space, $\vec{x}$, by varying each stellar parameter by the minimum grid step and observing the change in flux. In order to obtain the diagonal matrix, $\mathbf{C}_{f}$, we estimate the uncertainty in flux across the wavelength range by taking the publicly available\footnote{available from \url{https://www.cosmos.esa.int/web/gaia/iow\_20211223}} ESA/Gaia/DPAC signal-to-noise ratio as a function of wavelength for Gaia DR3 mean spectra. This provides us with the uncertainty in flux as a function of wavelength and G-band magnitude. We have additionally considered the impact of varying temperature on the signal-to-noise ratio (SNR), however the impact is not significant. The maximum SNR difference between temperatures is $\sim 20\%$ and thus we see that magnitude is the most important parameter in defining the SNR as a function of wavelength. Equation~\eqref{eq:Fisher} returns the maximum information available in each stellar parameter at a given position, $\vec{x}$, in the stellar parameter space ($T_{\rm eff}$, log$g$, [Fe/H]).

\begin{figure}
	\includegraphics[width=\columnwidth]{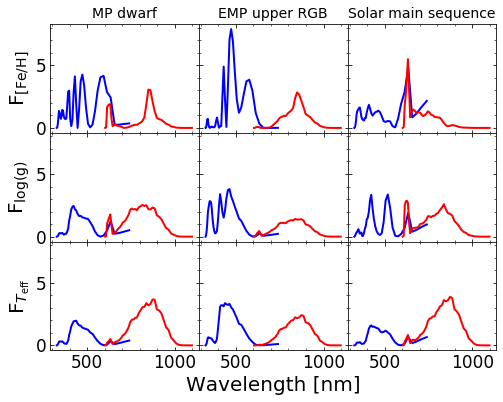}
    \caption{The Fisher information available at each wavelength of the BP and RP spectra normalised by the mean Fisher information value. The top row indicates metallicity information, the middle row shows surface gravity information and the bottom row shows effective temperature information. The blue line indicates the BP spectra information, while the red line shows the RP spectra information. These are shown for the same three example stars as shown and described in Figure~\ref{fig:SynthVsObs} at a fixed magnitude $G = 16$}.
    \label{fig:InfoVsWave}
\end{figure}

$\mathbf{F}(\vec{x},\lambda)$, in Equation~\eqref{eq:Fisher}, allows us to analyse the Fisher information in each stellar parameters at the position $\vec{x}$ as a function of wavelength as shown in Figure~\ref{fig:InfoVsWave}. We can see clear peaks in metallicity information in Figure~\ref{fig:InfoVsWave} that correlate to large or multiple absorption lines in Figure~\ref{fig:SynthVsObs} such as the Calcium triplet and Balmer series. We do see in some panels a notably higher information at the overlap between BP and RP spectra. This is due to a combination of factors: the largest signal-to-noise ratio is observed for many stellar types at $\sim$ 630 nm, with a significant decrease at larger wavelengths, we additionally see the largest transmission across BP and RP spectra at $\sim$ 630 nm in RP spectra. These all combine to amplify any information in this region. While the information available in surface gravity varies as a function of wavelength for different stellar types, the temperature information is largely confined to the continuum of the spectrum.

Although we are able to assess the Fisher information as a function of wavelength, it is most useful to assess the total information that we can extract from BP/RP spectra about a given star's stellar parameters. We can see from equation~\eqref{eq:Fisher} that the total Fisher information matrix is the sum of the Fisher matrices across the wavelength range and this is the best-case co-variance matrix for our stellar parameters at a given position in the stellar parameter space that is achievable given our model BP/RP grids. In summary, our application of the Fisher equation provides uncertainties such as $\sigma(T_{\rm{eff}})$, $\sigma(\rm{log}(g))$ and $\sigma(\rm{[Fe/H]})$.

In addition, given the co-variance matrix, we can also calculate the correlation between parameters, $\rho_{x,y} = \frac{\sigma_{x,y}}{\sigma_{x} \sigma_{y}}$, in order to assess how external constraints on stellar parameters can help to reduce the uncertainties determined using our BP/RP spectra.

\subsection{Effective temperature and surface gravity priors}
\label{sec:Priors}
The methods used in Section~\ref{sec:Fisher} give us both the best-case uncertainty on each stellar parameter across the parameter space and also the correlation between these parameters. We can first use these uncertainties to assess the feasibility of stellar parameter measurements, but we can additionally use the correlations between parameters to allow us to identify in what parameter space priors become most advantageous.

Although BP/RP spectra will prove invaluable for the determination of $T_{\rm eff}$ and log($g$), we note that Gaia's precise broadband photometry and parallaxes will allow us to obtain additional information on their probability distribution \citep{Anders2022}. These prior uncertainties will allow us to add extra information into equation~\eqref{eq:Fisher}, reducing the resultant uncertainty in metallicity.

\citet{Anders2022} found using Gaia EDR3 they were able to constrain $T_{\rm eff}$ to $\sim 140$ K and log($g$) to $\sim$ 0.2 dex. The ability to constrain these parameters without analysing BP/RP spectra, not only allows us to study false positive rates for specific effective temperatures and surface gravities as these can be independently identified, we can also use these to reduce the metallicity uncertainties and hence false positive rates.

\section{Fisher matrix results}
\subsection{Parameter uncertainties}
\label{sec: Uncertainties}

\begin{figure}
	\includegraphics[width=\columnwidth]{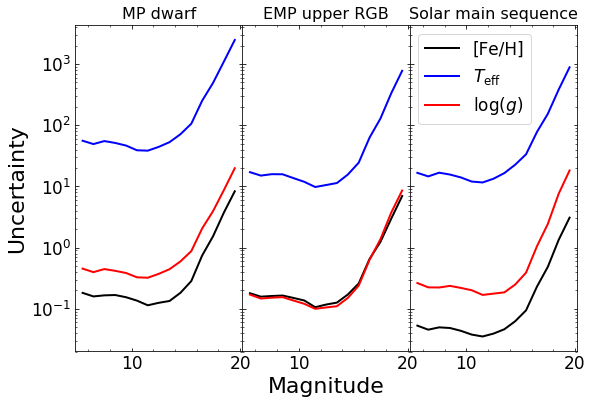}
    \caption{Parameter uncertainty as a function of G-band magnitude for the three example stars in Figure~\ref{fig:SynthVsObs} and Figure~\ref{fig:InfoVsWave}. The blue, red and black lines represent $\sigma(T_{\rm{eff}}[\rm{K}])$, $\sigma(\rm{log}(g)[\rm{dex}])$ and $\sigma(\rm{[Fe/H]}[\rm{dex}])$ respectively.}
    \label{fig:UncVsMag}
\end{figure}

\begin{figure}
	\includegraphics[width=\columnwidth]{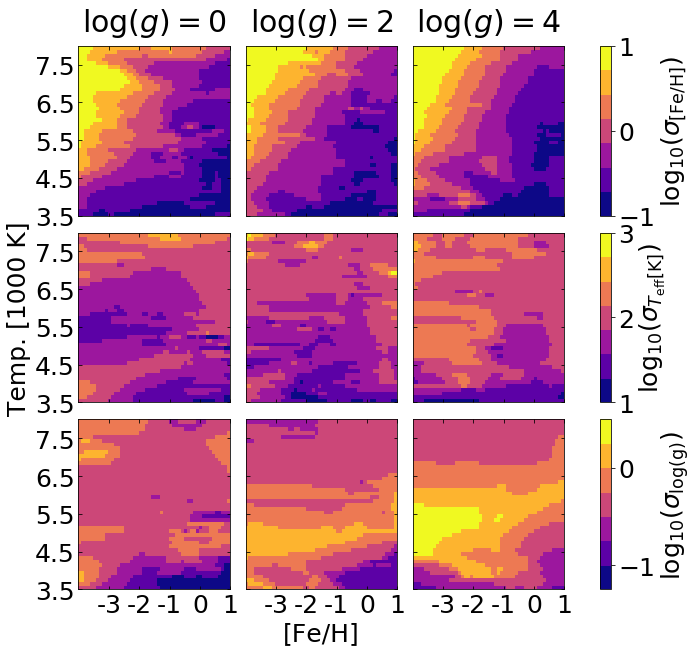}
    \caption{A grid of plots indicating the uncertainty of each parameter for different stellar parameters at magnitude $G = 16$. The x and y axes show increasing metallicity and effective temperature respectively, while each column shows a different surface gravity. Each row shares a common color-bar, the upper row's color-bar indicates the metallicity uncertainty, the middle row color-bar indicates effective temperature uncertainty and the lower row shows surface gravity uncertainty.}
    \label{fig:NoNormGrid}
\end{figure}

In this subsection we analyse the results of our Fisher analysis while neglecting any priors that may be available. This allows us to understand the information that is implicit to BP/RP spectra. We firstly assess the change in parameter uncertainty as a function of magnitude in Figure~\ref{fig:UncVsMag} for our example stellar types. Figure~\ref{fig:UncVsMag} shows stellar parameter uncertainties of the same order in the range of magnitudes $5 \leqslant G \lesssim 16$. We see a slight increase in uncertainty at magnitudes brighter than $G = 11$, due to a gradual increase in the signal-to-noise ratio at the brightest magnitudes. For $G > 16$ we see an exponential growth in all parameter uncertainties. Given this pivot point, for the following analysis we use a magnitude of $G = 16$ as it gives us the upper-bound of stable parameter uncertainties before they start to exponentially grow with magnitude. 

Figure~\ref{fig:NoNormGrid} shows the best-case uncertainties we find on each stellar parameter for a range of input stellar parameters. We see little change in effective temperature and metallicity uncertainties across changing surface gravity. We clearly see that metallicity uncertainty is a function of both metallicity and effective temperature. Increasing temperature acts to decrease the relative depth of absorption lines, which provide information on the metallicity of a star, thus increasing the metallicity uncertainty. Decreasing metallicity has the same effect as increasing temperature, and hence we see a gradient in metallicity uncertainty across the effective temperature and metallicity parameter space, as seen in the top row of grids in Figure~\ref{fig:NoNormGrid}. 

We can see that although there are mild structures and trends in effective temperature uncertainty across the parameter space, the clear result is that using Gaia BP/RP spectra we have an effective temperature resolution of order 100 K. 

The clear result for surface gravity measurements is an increasing uncertainty for increasing surface gravity. This however appears to occur most significantly in a effective temperature range of 4500 K $< T_{\rm eff} < $ 6500 K. This creates a change from a resolution of $\sim$ 0.3 dex for log$(g) = 0$, up to a resolution of $\sim$ 3 dex for log$(g) = 4$. Outside of this region of significant variation with surface gravity, we see a resolution of order 0.3 dex for $T_{\rm eff} > $ 6500 K, down to a best resolution of 0.1 dex for some regions of the parameter space with $T_{\rm eff} <$ 4500 K. We can therefore clearly see that for many intrinsic stellar parameters we shall be able to discriminate between dwarfs and giants from surface gravity measurements using Gaia BP/RP spectra.

We can see from Figure~\ref{fig:NoNormGrid} that for low temperature stars (< 4500 K) and for stars with metallicity [Fe/H] $> -1$ we are able to constrain metallicity uncertainties to of order $\sigma_{\rm [Fe/H]} \sim$ 0.1 dex. This uncertainty rapidly grows as we move to higher temperatures and to metallicities below [Fe/H] $< -1$, with uncertainties of order 1 dex, reaching a maximum uncertainty larger than 10 dex as we reach temperatures of 7500 K.

\subsection{Parameter correlations}
\label{sec:correlations}

\begin{figure}
	\includegraphics[width=\columnwidth]{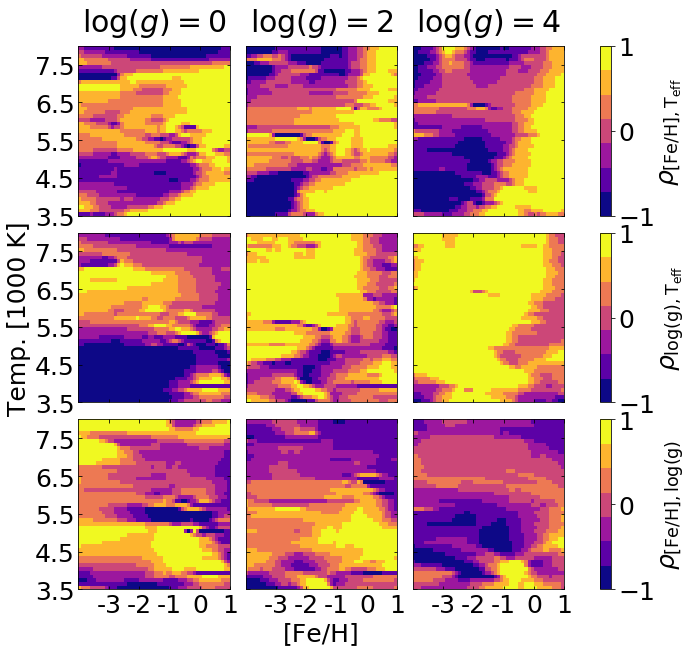}
    \caption{A grid of plots indicating the correlations between each of the stellar parameters we investigate. The x and y axes show increasing metallicity and effective temperature respectively, while each column shows a different surface gravity. The upper row has a common color-bar indicating the correlation between metallicity and effective temperature, the middle line color-bar indicates the correlation between surface gravity and effective temperature and the lower row shows surface the correlation between metallicity and surface gravity.}
    \label{fig:Correlation Grid}
\end{figure}

The uncertainties shown in Figure~\ref{fig:NoNormGrid} could potentially be reduced by including temperature and surface gravity priors, evaluated using Gaia's precise photometry and parallaxes, in equation~\eqref{eq:Fisher}. We must first however assess whether there are correlations between parameters that would allow for the exploitation of such priors. This is done using methods discussed in section~\ref{sec:Fisher}. Figure~\ref{fig:Correlation Grid} shows the results of this analysis, indicating the correlations between all free stellar parameters in our BP/RP mock spectra grid.

We can clearly see in Figure~\ref{fig:Correlation Grid} that metallicity is largely correlated with both temperature and surface gravity. We expect that in regions of Figure~\ref{fig:NoNormGrid} where temperature uncertainty is greater than 140 K, and where Figure~\ref{fig:Correlation Grid} shows temperature and metallicity are strongly correlated, that the metallicity uncertainty will be further constrained with the use of a temperature prior. This is similarly true for surface gravity priors. 

We can see from Figure~\ref{fig:NoNormGrid} that we have effective temperature resolution $\sigma_{T_{\rm eff}} \sim$ 100 K across most of the stellar parameter space. There are however small regions, for instance at log$(g)$ = 4, for metallicities [Fe/H] $< -2$ and temperatures 4500 K $< T_{\rm eff} < $ 6000 K where we see temperature uncertainties greater than the 140 K prior. Within this region we see a strong anti-correlation between temperature and metallicity uncertainties, and therefore we can expect a small reduction in metallicity in this region when we account for temperature priors. We expect no contribution from a temperature prior in the remainder of the stellar parameter space. 

In addition, we can see from Figure~\ref{fig:NoNormGrid} that the uncertainty in surface gravity is larger than the prior of 0.2 dex across almost the entire parameter space. Figure~\ref{fig:Correlation Grid} indicates largely strong correlations between surface gravity and metallicity across the parameter space. We therefore expect any improvement in metallicity uncertainty acquired due to the inclusion of priors to be driven by surface gravity priors. We see the lowest surface gravity uncertainties around $T_{\rm eff} \sim 3500 \: \rm{K}$ and we therefore expect minimal metallicity uncertainty improvement with a surface gravity prior at this temperature. We additionally see at log$(g)$ = 2 and 4, for temperatures above 6000 K respectively, there is minimal correlation between metallicity and surface gravity. Further to this, when regions move from positive to negative correlation, we see no correlation at these boundaries. At all of these regions of low or no correlation between metallicity and surface gravity, we expect to see minimal or no improvement in metallicity uncertainty when we include a surface gravity prior. 

Due to the surprising nature of some of these changes in correlations between parameters from positive to negative, we have additionally used a basic minimisation of the reduced chi-squared statistic model to assess the correlation between parameters. We simulate observed spectra by adding random noise to our model spectra based on the SNR data used in section~\ref{sec:Fisher} and by iterating over all model spectra to minimise the reduced chi-squared statistic we find the best-fit stellar parameters. We repeat this multiple times to find the spread in best-fit stellar parameters and hence estimate the uncertainty and correlation between parameters. We ultimately find correlations and uncertainties consistent with those seen in Figure~\ref{fig:Correlation Grid} and~\ref{fig:NoNormGrid} using Fisher information matrices.

\subsection{Metallicity uncertainty with priors}
\label{Uncertainty with priors}
\begin{figure*}
	\includegraphics[width=\textwidth]{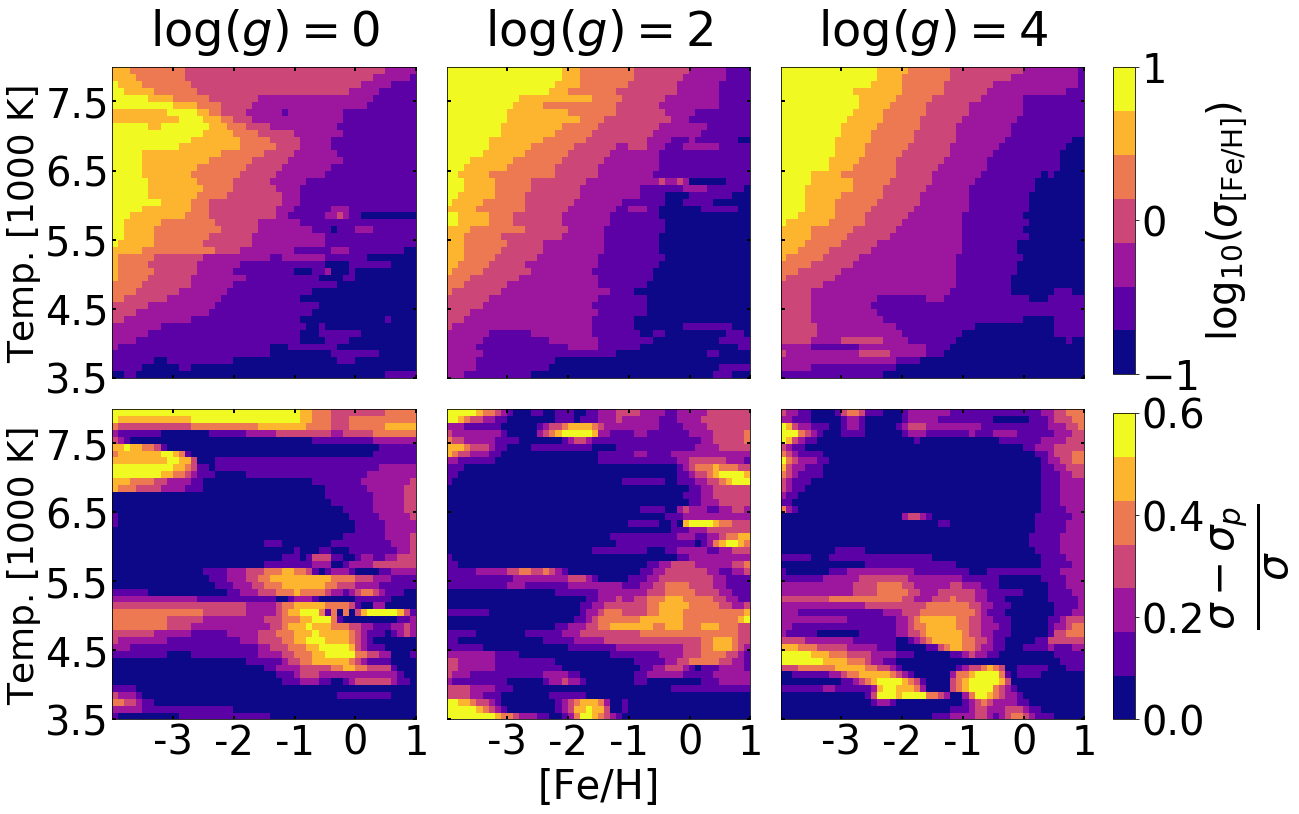}
    \caption{The top row shows metallicity uncertainty plots, for a star of magnitude $G = 16$, when we include effective temperature (140 K) and surface gravity (0.2 dex) priors. The bottom row indicates the fractional improvement from the original metallicity uncertainty, $\sigma$, when we include priors to produce $\sigma_{p}$. The x and y axes show increasing metallicity and effective temperature respectively, while each column shows a different surface gravity. Each row shares a common color-bar indicating the metallicity uncertainty and fractional improvement in metallicity uncertainty for the upper and lower rows respectively.}
    \label{fig:Uncertainty with prior}
\end{figure*}

Figure~\ref{fig:Uncertainty with prior} shows the metallicity uncertainty after we include temperature and surface gravity priors and additionally the fractional improvement in the metallicity uncertainty provided by including these priors. Figure~\ref{fig:Uncertainty with prior} confirms the hypotheses made in Section~\ref{sec:correlations} -- we see a notable reduction in metallicity uncertainty that mostly traces strong correlations in the surface gravity correlation plots in Figure~\ref{fig:Correlation Grid}. We see little fractional improvement in the upper regions of the log$(g) = 2$ and 4 and additionally, along the boundaries of regions of positive and negative correlations, where we have minimal correlation between metallicity and surface gravity.

We can see in the lower row of Figure~\ref{fig:Uncertainty with prior} that the inclusion of surface gravity and effective temperature priors can improve metallicity uncertainties by up to 60$\%$. We now find that across all surface gravities, for the half of the plot that has either low effective temperature, high metallicity or both has a metallicity uncertainty of less than 0.3 dex. 

It is clear from these plots that for large regions of the stellar parameter space, we are able to reduce the metallicity uncertainties to low enough values as to confidently constrain the metallicity of a star using BP/RP spectra.

\section{Identifying EMP stars with Gaia BP/RP }
\subsection{Mock observed population}
\label{sec:Mock Population}
In order to assess the effects these uncertainties may have on the observed metallicity distribution one may observe using Gaia BP/RP, we must start wish some initial distribution of stellar parameters. We chose to use Gaia Object Generator (GOG) to produce our intrinsic distribution \citep{Antiche}. GOG is a simulation tool based on the Gaia Universe model to produce a catalogue of the potential observable stellar population (with magnitudes brighter than $G = 20$). The assumed metallicity distribution function of the model is composed of individual metallicity distributions for each stellar population - the disc, halo and bar. Each population is represented by a normal distribution centred on a mean value and truncated at 5 sigma, as such the minimum metallicity in the mock population is [Fe/H] $= -4$ \footnote{The values of the mean and sigma for each metallicity distribution can be found at \url{https://gea.esac.esa.int/archive/documentation/GEDR3/Data_processing/chap_simulated/sec_cu2UM/ssec_cu2starsgal.html}}. 

Some of the data included within this catalogue are the stellar parameters that we are interested in: magnitude, effective temperature, surface gravity and metallicity. This therefore gives us an ideal intrinsic stellar population for which we are able to estimate the best-case uncertainties of individual stars.

We take this intrinsic population and focus primarily on their metallicites. We focus on three "low-metallicity" regimes, [Fe/H] < -2, -2.5 and -3 (0.5$\%$, 0.07$\%$ and 0.004$\%$ of the total GOG population respectively). We take this intrinsic metallicity population and randomly re-draw each metallicity from a Gaussian distribution centred on its intrinsic metallicity with a standard deviation equal to the uncertainty derived from Section~\ref{sec:Fisher}. Doing so produces a mock observed metallicity distribution. We use this to test the numbers of "high metallicity" stars that are shifted into the low-metallicity regime due to the uncertainty in their measurement using Gaia BP/RP spectra. This finally allows us to assess the false positive rates (FPR) for detections of stars with low metallicities for the three low-metallicity regimes. We produce this analysis using the uncertainties from Section~\ref{Uncertainty with priors} including the assumed effective temperature and surface gravity priors. Ignoring these priors has a minimal impact on our results.

We note that our assumption that the metallicity distribution is a Gaussian centred on the intrinsic metallicity becomes poor for large uncertainties. Through the use of our reduced chi-squared minimisation method (discussed in section~\ref{sec:correlations}) we analysed the biases in the predicted metallicity distribution. We find that for Fisher information uncertainties greater than roughly 1 dex, the assumption of a Gaussian distribution breaks down. The distribution is instead biased towards higher metallicities, therefore, stars with non-Gaussian distributions are more likely to be scattered away from "low-metallicity" regions, and hence are less likely to pollute our low-metallicity detections. We therefore conclude that our inability to incorporate these distributions into our predictions will act to increase our FPR, meaning the results from this analysis may be slightly over-estimating the FPR. 

\subsection{Simulating EMP detections}
Although GOG gives us detailed information on a sample of 523 million simulated Gaia observed sources, due to data processing constraints, we choose to reduce to a sub-sample of 215 million stars with simulated effective temperatures, surface gravities and metallicities. 

Following the analysis described in Section~\ref{sec:Mock Population} we first find the FPR for a range of magnitude bins in Table~\ref{tab:FPR}. Given Figure~\ref{fig:UncVsMag} indicates that metallicity uncertainty exponentially grows for magnitudes $G > 16$, we would expect the FPR to have a similar trend, which can be observed in Table~\ref{tab:FPR} where we see the largest growth in FPR at $G \sim 16$. These results indicate that for stars of magnitude $G > 16$, for the metallicity regimes considered, many intrinsically metal-poor stars are lost within crowds of false positive detections. However, we note that for stars of magnitude brighter than $G = 16$, for the metallicity regime [Fe/H] < -2, at least 3 in every 4 low-metallicity detections are true positives. This rate is at least 3 in every 5 for the metallicity regime [Fe/H] < -2.5 at magnitude $G < 15$, and roughly 1 in every 2 for [Fe/H] < -3 at magnitudes $G < 14$. We would expect the FPR for [Fe/H] < -3 to further improve up to $G = 11$, however, small number statistics mean we cannot confidently comment at such bright magnitudes. 

\begin{table}
\caption{Table indicating the false-positive rate of EMP star detections for a range of low metallicity regimes and at varying stellar magnitudes. The true-positive rate can be found by taking ($1 -$ FPR) reported in the table. }
\label{tab:FPR}
 \begin{tabular}{lccc}
  \hline
  Mag. & & False Positive Rate & \\
   & [Fe/H] < -2 & [Fe/H] < -2.5 & [Fe/H] < -3 \\
  \hline
  $20 \geq G > 19$ & 0.97 & 1.00 & 1.00 \\
  $19 \geq G > 18$ & 0.95 & 0.99 & 1.00\\
  $18 \geq G > 17$ & 0.87 & 0.96 & 1.00\\
  $17 \geq G > 16$ & 0.66 & 0.88 & 0.98\\
  $16 \geq G > 15$ & 0.31 & 0.61 & 0.90\\
  $15 \geq G > 14$ & 0.20 & 0.42 & 0.73\\
  $14 \geq G > 13$ & 0.14 & 0.35 & 0.69\\
  \hline
 \end{tabular}
\end{table}

\begin{figure*}
  \centering
  \caption{A Kiel diagram indicating the stellar parameter distribution of GOG stars with simulated BP/RP observed metallicities [Fe/H] < -2. The population of "low-metallicity" stars is binned into effective temperature and surface gravity bins and the FPR within each bin is indicated by the colorbar. Histograms on a logarithmic scale, located on the x and y axes, indicate the distribution of true positive (hatched blue) and false positive (filled grey) detections. The FPR as a function of each stellar parameter is additionally over-plot in red.}
     \begin{subfigure}[b]{0.48\linewidth}
         \centering
         \includegraphics[width=0.9\hsize]{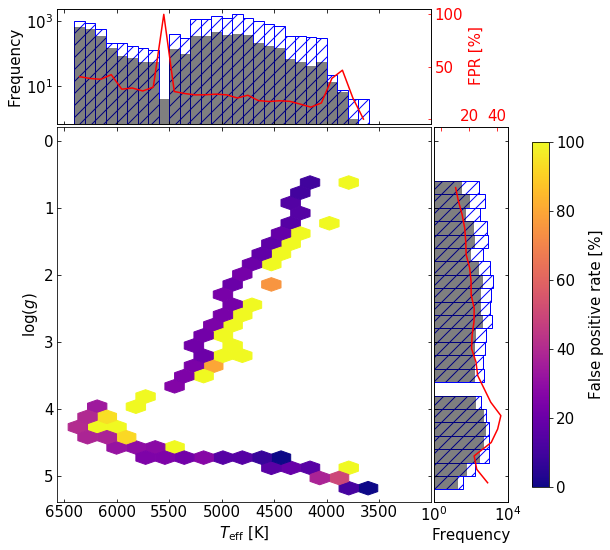}
         \caption{This Kiel diagram only includes stars with intrinsic magnitudes brighter than $G = 16$.}
         \label{fig:Intrinsic Paramaters 16}
     \end{subfigure}
     \hfill
     \begin{subfigure}[b]{0.48\linewidth}
         \centering
         \includegraphics[width=0.9\hsize]{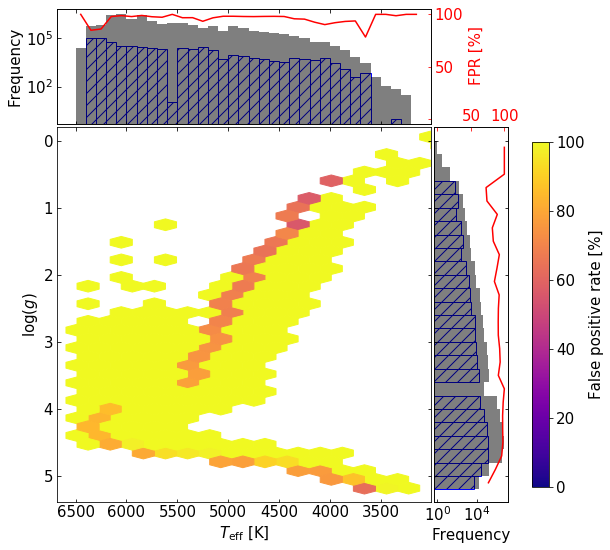}
         \caption{This Kiel diagram only includes stars with intrinsic magnitudes dimmer than $G = 16$.}
         \label{fig:Intrinsic Paramaters greater}
     \end{subfigure}
\end{figure*}

In order to best understand the driving mechanisms behind false positive detections, we analysed the stellar parameter distribution for "low-metallicity" stars. Figure~\ref{fig:Intrinsic Paramaters 16} and~\ref{fig:Intrinsic Paramaters greater} show the FPR within a set of bins for stars with observed metallicity [Fe/H] < -2. They additionally indicate the distribution of false and true positive detections across each dimension of the parameter space in the form of histograms and the FPR is overlaid onto these plots. These allow us to best understand the driving processes behind false-positive detections of metal-poor stars using Gaia BP/RP spectra.

Figure~\ref{fig:Intrinsic Paramaters 16} shows the population of stars with magnitude brighter than $G = 16$. It clearly indicates that the source of true positive detections fall mostly along the isochrones where we would expect to find low-metallicity stars, while at cooler temperatures away from these isochrones, where we know MP stars are rarer, we find a near 100$\%$ FPR. Further analysis of the metallicity distribution finds that the source of false positive detections at such bright magnitudes are in fact stars that have metallicities near to our "low-metallicity" boundary. We find bins that are largely false-positive detections in Figure~\ref{fig:Intrinsic Paramaters 16} have a median intrinsic metallicity of [Fe/H] $\lesssim -1.5$. 

Therefore our harsh metallicity cut-off makes our FPR appear more pessimistic than reality, as a false positive detection for a [Fe/H] < -2 star has a high probability of still being a "low-metallicity" star, just not at the extreme of [Fe/H] < -2. This is emphasised in Table~\ref{tab:FPR}, where we see large increases in FPR between the three "low-metallicity" regimes at bright magnitudes, indicating that when we observe these large increases, the increase in FPR is driven by stars with metallicities between such metallicity regimes. This can therefore further act to support a [Fe/H] < -3 detection using Gaia BP/RP spectra, as it appears even if any such detection is a false positive, there is a reasonably good change the intrinsic metallicity will still fall between -2 < [Fe/H] < -3.

Figure~\ref{fig:Intrinsic Paramaters 16} also indicates that there is an increase in the FPR at temperatures between 3500 K $< T_{\rm eff} < 4000$ K and additionally a peak at $T_{\rm eff} = 5550$ K. These features are repeatedly observed at a variety of magnitudes, and although the small number of stars in this population means that any such local peak in FPR doesn't significantly effect the global FPR, we advise that any "low-metallicity" detections within these temperature ranges be treated with caution. We additionally see a mild trend, outside of the offending temperature range, of increasing FPR with increasing temperature. This is expected given the increasing metallicity uncertainty with temperature discussed in Section~\ref{sec: Uncertainties}, however the minimal overall increase within the temperature ranges present in the data shown in Figure~\ref{fig:Intrinsic Paramaters 16} means there is no need to place any selection criteria on low-metallicity stellar detections based on this trend. The FPR appears to increase with increasing surface gravity, however any such increase is minimal enough that we still have a relatively low FPR across our surface gravity range and hence readers do not need to target specific surface gravity ranges in their "low-metallicity" identification strategy. Our results ultimately find that if exterior measurements on effective temperature and surface gravity allow a star to be confined to one of the bins in Figure~\ref{fig:Intrinsic Paramaters 16}, there are regions that allow for a near 100$\%$ success-rate for metal-poor detections. Constraining stars to such regions, and consequently finding stars in regions with large FPRs, will allow the community to hand-pick from their population of metal-poor detections, to create a population of high-confidence metal-poor detections for follow-up high-resolution observations.

What is clear from this analysis is that [Fe/H] < -2 detections for magnitudes brighter than $G = 16$ are not polluted by stars of high metallicity that have exceedingly large uncertainties. This situation becomes the driver for false positive detections at dimmer magnitudes, $G > 16$. The exponential growth of uncertainties at larger magnitudes means that we collect false positive detections from a larger range of stellar parameters, due to their large uncertainties, as seen in Figure~\ref{fig:Intrinsic Paramaters greater}. We investigate whether we have stellar parameters for which we can accurately identify metal-poor stars ([Fe/H] < -2) when considering the magnitude range dimmer than $G = 16$ in Figure~\ref{fig:Intrinsic Paramaters  greater}. 

We initially apply a cut to the hottest stars and white dwarfs, as modelling of their isochrones are notoriously challenging and hence intrinsic metallicity estimates are spurious. All such stars had a 100$\%$ FPR, and therefore are not of interest. We encourage anyone using Gaia BP/RP spectra to identify metal-poor stars to adopt a similar strategy as these stellar parameters are not observed for magnitudes brighter than $G = 16$ in GOG, and are exclusively false-positive detections for magnitudes $G > 16$. We can see that when observing magnitude $G > 16$ stars, their large uncertainties mean that we have are heavily polluted with false positive detections across the entire temperature and surface gravity range considered. The FPR is $\sim 100 \%$ for the vast majority of stellar parameters, however, along the isochrones where we ordinarily expect to find metal-poor stars, the FPR can fall to values around 60$\%$. This may allow for metal-poor detections of reasonable confidence, especially when either temperature or surface gravity has a more precise prior than we consider or additionally for magnitudes that lie close to $G = 16$.

\subsection{Comparison with observational metal-poor surveys}
There are several photometric surveys aiming to successfully identify metal-poor stars as tracers of formation and evolution of the Galactic halo. One of the pioneering examples was the \ion{Ca}{II} H\&K objective-prism survey by \citet{bee85} that reported more than a hundred bright very metal-poor (VMP) candidates ($\rm [Fe/H]$<-2). This represented a leap in Galactic archaeology and showed a promising methodology to identify stars from this rare class. Then, a very important contribution was provided by the Hamburg/ESO survey \citep{Wisotzki2000,chis01} also based on the Ca II resonance lines around 395\,nm. Thanks to this objective-prism survey, \citet{fre06} identified 1718 metal-poor candidates with $\rm B_{mag}<13$, and confirmed 174, 98, 23 to be more metal-poor than [Fe/H]=-2.0, -2.5, and -3.0 respectively. Some of the most metal-poor stars ever known were primarily identified by the Hamburg/ESO survey \citep{christlieb02, fre05}.
\
A few years later, a dedicated survey, the Skymapper \citep{keller07}, used narrow-band filter photometry to identify metal-poor candidates. With this technique a larger number of fainter objects could be targeted and the amount of metal-poor candidates increased drastically. More recently, thanks to the Skymapper narrow-band filter, \citet{dacosta2019} reported from more than 2600 metal-poor candidates followed-up with medium-resolution spectroscopy, 93\%, 41\% and 18\% success ratios for stars with $\rm [Fe/H]$<-2.0, -2.75, and -3.0 respectively.
Some other relevant searches were completed as part of the Sloan Digital Sky Survey \citep[SDSS,][]{york00} such as the TOPoS survey \citep{caffau13} or follow-up on EMP stars by \citet{agu16}. Also by means of the large Sky Area Multi-Object Fiber Spectroscopic Telescope \citep[LAMOST,][]{Deng12} like the 10,000 VMP stars reported by \citet{li18}.
Moreover, in the northern hemisphere, the Pristine survey \citep{St17} used even narrower photometry around the H\&K area and provided high-quality photometric metallicities for around 2 million metal-poor stars. \citet{agu19b} reported after a medium-resolution follow-up of more than a thousand metal-poor candidates, success ratios of 88\%, 56\%, and 23\% for stars with $\rm [Fe/H]$<-2.0, -2.5, and -3.0 respectively. 
Additionally, a multi-filter photometric survey, J-Plus \citep{whitten2019}, is also identifying metal-poor stars based on their colours. In a recent paper by \citet{galarza22}, the authors observed 177 metal-poor candidates selected with J-Plus photometry and confirmed 64\% of them to be more metal-poor than [Fe/H]=-2.5.
Finally, the Best and the Brightest survey \citep{Sc14} using a combination of public photometric data in different bands identified efficiently bright ($V_{mag}<14)$ metal-poor candidates (32\%, and 4\% for [Fe/H]<-2.0, and -3.0 respectively). According to \citet{Li21} when this multi-band selection is combined with Gaia EDR3 astrometry \citep{gaiaedr3} the efficiency grows significantly (76\%, 28\%, and 4\% for [Fe/H]$ \leq  - 2.0, \leq - 2.5$ and $ \leq - 3.0$, respectively).

We find that when we apply a similar magnitude cut of $G < 14$ to the GOG data, we have a success rate of metal-poor identification of 86\%, 67\% and 44\% for [Fe/H]$ \leq  - 2.0, \leq - 2.5$ and $ \leq - 3.0$, respectively. These values are a significant improvement on the results from \citet{Li21} that uses only Gaia's astrometry, indicating that the inclusion of BP/RP spectra in such surveys has the potential to drastically improve the success-rate of metal-poor star detections. We additionally see that the success rate for the lowest metallicity regime ([Fe/H] < -3) offers a significant improvement on all other narrow-band photometric surveys indicating that Gaia BP/RP spectra have the potential of more accurately identifying populations of extremely-metal-poor stars than any previous metal-poor survey.

\section{Chemistry with Gaia BP/RP}
\subsection{Carbon Sensitivity}
\label{sec:Carbon Analysis}
Carbon enrichment in halo stars is a crucial issue in Galactic archaeology. The fraction of metal-poor stars that show $\rm [C/Fe]$>0.7 increases dramatically with decreasing metallicity \citep[see e.g.,][and references therein]{Cohen_2005,placco14,bonifacio15, yoon18,arentsen21}. There are a number of explanations invoking both extrinsic \citep[see e.g.,][]{herwig05} and intrinsic \citep[see e.g.,][]{umeda03,Ishigaki14} processes for the creation of carbon-enriched metal-poor (CEMP) stars. It is crucial to be able to pre-select metal-poor stars that are carbon rich.

\begin{figure*}
	\includegraphics[width=\textwidth]{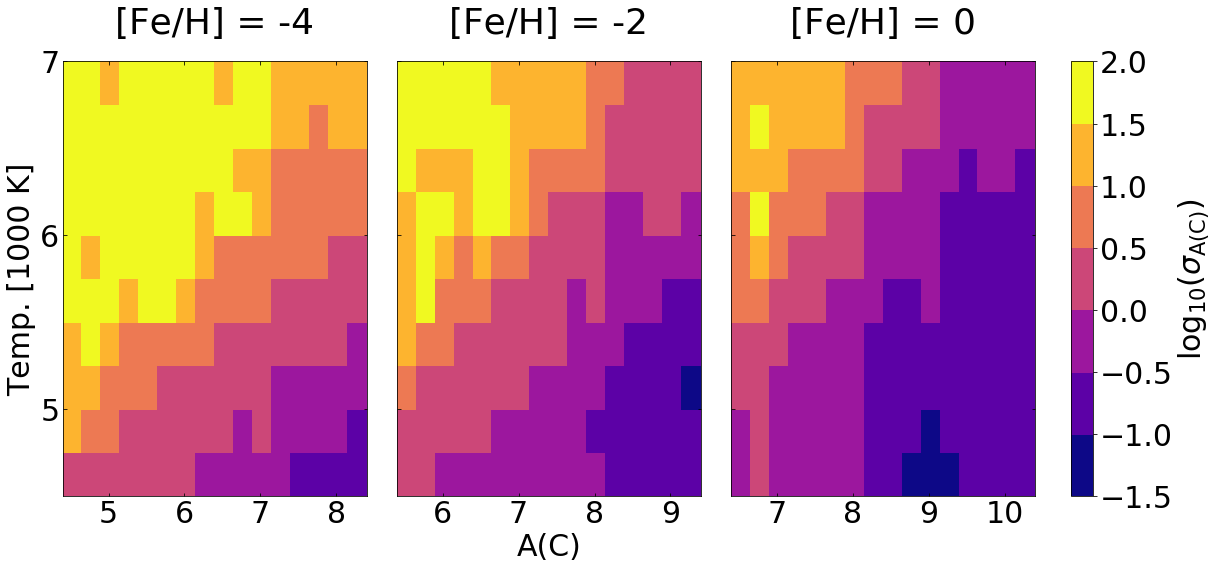}
    \caption{A row of Carbon abundance uncertainty plots, for a star of magnitude $G = 16$. The x and y axes show increasing absolute Carbon abundance and effective temperature respectively, while each column shows a different metallicity. The plots share a common color-bar indicating the absolute Carbon abundance uncertainty.}
    \label{fig:Carbon Uncertainty}
\end{figure*}

We investigate the potential for Gaia BP/RP spectra to resolve absolute carbon abundance, A(C), using the synthetic spectra grid, discussed in Section~\ref{sec:Carbon Grid}. Although these grids supply spectra that only extend from $355-1149$ nm, thus missing the bluest region of the BP spectrum, due to low transmission in this region, we expect this to have a minimal impact on the accuracy of our mock BP/RP spectra analysis. The synthetic spectra library includes Carbon as a free parameter and therefore following the same steps discussed in Sections~\ref{sec:BP RP Simulation} and~\ref{sec:Analysis} we can evaluate the potential resolution of Carbon abundance measurements using BP/RP spectra as well as the resolution on all other stellar parameters discussed in this paper. 

We initially fix the Carbon abundance at [C/Fe] = 0, in order to compare our results from the Carbon grid to the results from the grid used throughout the rest of this work (discussed in Section~\ref{sec:PHX}). For the stellar parameters for which the two grids overlap, we find consistent results regarding their uncertainties and correlation, confirming our results are invariant with the spectral library we use. 

Following this we focus on the absolute Carbon abundance, $\rm A(C) = [C/Fe] + [Fe/H] + 8.39$\footnote{The solar carbon abundance considered here is the one reported by \citet{asplund05}, A(C)$_{\odot}=8.39$.}, where the uncertainty in absolute Carbon abundance can be found by combining $\sigma_{\rm[C/Fe]}$ and $\sigma_{\rm[Fe/H]}$ in quadrature. Figure~\ref{fig:Carbon Uncertainty} shows the results of the absolute Carbon abundance analysis indicating the absolute Carbon abundance uncertainty as a function of metallicity and effective temperature on the x and y axes respectively, with each column showing a different metallicity. The analysis was completed for a star with fixed surface gravity log$(g) = 2$ as we found that absolute Carbon abundance uncertainty was not a function of the stellar parameter surface gravity. The results are for a fixed magnitude  of $G = 16$. 

Figure~\ref{fig:Carbon Uncertainty} shows an increasing absolute Carbon uncertainty as a function of increasing temperature and decreasing absolute Carbon abundance. Therefore, in order to place a reasonable constraint on Carbon abundance of $\sigma_{\rm A(C)} < 0.5$ the observed star must have an intrinsic absolute Carbon abundance A(C) $> 8$ and must have a temperature below 5000 K, 6000 K and 6500 K for metallicities [Fe/H] = -4, -2 and 0 respectively. 

We use a criterion of [C/Fe] > 0.7 for CEMP stars. Given this criterion, we can see that this equates to A(C) > 6.1, 7.1 and 8.1 for the three metallicity regimes shown in Figure~\ref{fig:Carbon Uncertainty}. If we place a temperature constraint of $T_{\rm eff} < 6000$ K for CEMP stars, we can see that for metallicities [Fe/H] = -2 and 0, we have an absolute Carbon abundance uncertainty $\sigma_{\rm A(C)} < 1$. We obtain a similar uncertainty for metallicity [Fe/H] = -4 when we constrain effective temperatures to below 5000 K. These uncertainties are large enough that we may not be able to confidently identify individual CEMP stars with such stellar parameters, especially when we consider false positive detections. However, these reasonable uncertainties for CEMP stars mean that the false negative rate will be low for CEMP stars. Therefore Gaia BP/RP spectra will be crucial for identifying candidate CEMP stars for high-resolution spectroscopic follow-up. Although the population identified as CEMP stars using Gaia BP/RP spectra will have some level of false positive pollution, it will contain the vast majority of true CEMP stars and therefore this population will be essential for follow-up studies of such stars. 

While discovering Carbon-enhanced, hyper metal-poor stars ([Fe/H] $\sim$ -5) would be of huge interest and represent the frontier of the earliest chemical enrichment of the universe, using BP/RP from Gaia DR3 we unfortunately find it to be very unlikely. Our analysis extends down to [Fe/H] = -4.0, but by extrapolating our results in Figure~\ref{fig:Carbon Uncertainty} to lower metallicities, we expect that only very cool, extremely carbon-rich (A(C) $\sim$ 9 or 10) stars could possibly show Carbon features in hyper metal-poor stars. Nevertheless, we note that EMP stars ([Fe/H] $\sim$ -2, -3) with high carbon abundances play an important role in galactic archaeology and in that case Gaia BP/RP will help to identify some of them.

\subsection{\texorpdfstring{$\alpha$}{TEXT}-element abundance Sensitivity}
Elements created by fusing helium nuclei (O, Mg, Si, S, Ca, Ti) are so-called $\alpha$-elements and play an important role in Galactic Archaeology \citep[see e.g.,][]{shetrone03}. Pioneering works reporting the $\alpha$-excess in halo metal-poor stars \citep[see e.g.,][]{aller60,allerstein62} helped to understand the relation between the relative abundances of $\alpha$-elements with respect to iron and the star formation history of a system such as the Milky Way or its satellites. Since then, hundreds of authors have studied the large scale properties related to the chemistry of $\alpha$-elements \citep[see e.g.,][and references therein]{edvardsson93,NS97}. Moreover, significant efforts are still underway to try to understand the important role of $\alpha$-production in different cosmic environments. 
\begin{figure}
	\includegraphics[width=\columnwidth]{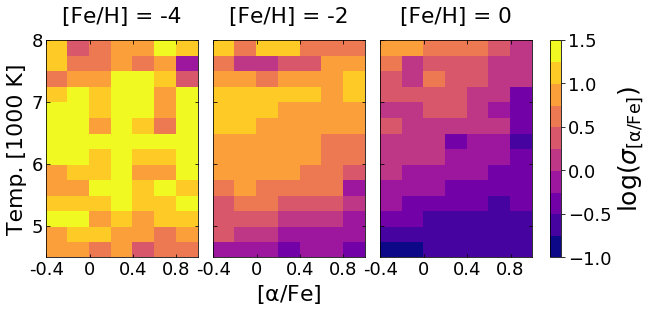}
    \caption{A plot of $\alpha$-element abundance for a star of magnitude $G = 16$. The horizontal and vertical axes show increasing $\alpha$-element abundance and effective temperature respectively. The colour-bar shows the $\alpha$-element abundance resolution.}
    \label{fig:Alpha Uncertainty}
\end{figure}

Throughout the analysis discussed in Section~\ref{sec:Analysis} we focused on the stellar parameters: effective temperature, surface gravity and metallicity. The stellar library discussed in Section~\ref{sec:PHX} however additionally includes $\alpha$ abundance as a free parameter. Similarly to Section~\ref{sec:Carbon Analysis} we repeat our original analysis while including $\alpha$-element abundance as a parameter. 

The results of this analysis can be seen in Figure~\ref{fig:Alpha Uncertainty} where we see that $\alpha$-element abundance uncertainty increases as a function of increasing effective temperature and additionally as a function of decreasing abundance, a feature that is common across all abundance uncertainty plots shown in this paper (Figures~\ref{fig:NoNormGrid}, \ref{fig:Uncertainty with prior} and~\ref{fig:Carbon Uncertainty}). We note that we only see an $\alpha$-element abundance uncertainty of $\sigma_{\rm [\alpha/Fe]} < 0.3$ dex for high metallicity, [Fe/H] = 0, and low temperature, $T_{\rm eff} \lesssim 5000$ K. Assuming $\alpha$-abundance varies between  $-0.4 < [\rm \alpha/Fe] < 1$ dex, a resolution that is of order 1 dex does not allow for any reasonable constraint on the $\alpha$-element abundance of the star. Therefore, for many stellar parameters we will be unable to use BP/RP spectra to constrain $\alpha$-element abundance, as suggested by \cite{Gavel+22}. Our results suggest that when temperature and metallicity measurements identify a star as being high metallicity ([Fe/H] = 0) and low temperature ($T_{\rm eff} \lesssim 5000$ K), we will likely have a low $\alpha$-element abundance uncertainty. This low uncertainty may allow us to identify populations of low or high $\alpha$-element abundance stars in this stellar parameter range, however, the large increase in uncertainty for any other stellar parameters means studying $\alpha$-element abundance variations across the stellar parameter space will potentially be problematic.

\section{Conclusions}

Gaia DR3 will see the release of over 100 million stellar spectra, producing a wealth of opportunities to study stellar parameters. However, the low resolution ($R \sim 50$) of the on-board BP and RP spectrographs were expected to lead to large uncertainties on measurements of stellar parameters such as metallicity. We produced mock BP/RP spectra using the known instrumentation of the spectrographs in order to study the feasibility of extracting the effective temperature, surface gravity and in particular metallicity of a star observed by Gaia. We produced a mock BP/RP spectra catalogue by transforming a synthetic spectra catalogue using the known transmission curve, spectral resolution and line-spread-function of the spectrographs. 

In order to obtain the best-case uncertainties on stellar parameter measurements, we analysed these mock spectra using Fisher information matrices. We found that for most intrinsic stellar parameters, for a magnitude $G = 16$ star, we can constrain the effective temperature to within 100 K, while surface gravity can often be constrained to a resolution that allows us to distinguish between dwarfs and giants. This however shows that, in combination with observed high correlations between stellar parameters, expected effective temperature and surface gravity priors of 140 K and 0.2 dex respectively will help to reduce metallicity uncertainties. Before introducing these priors, we found metallicity uncertainties that increase with decreasing metallicity and increasing temperature, ranging from of order 0.1 dex for low temperature, high metallicity stars, up to of order 10 dex for high temperature, low metallicity stars. Following the inclusion of temperature and surface gravity priors, we saw the same trends, but with improvements in metallicity uncertainty of up to 60$\%$ for some stellar parameters. 

We took simulated Gaia observed stellar parameters of 215 million stars and used these parameters to derive their associated metallicity uncertainty. Using this uncertainty we re-drew their metallicity to produce a mock-observed metallicity distribution. This then allowed us to probe the false positive rate of identifying a "low-metallicity" star using Gaia BP/RP spectra for a range of "low-metallicity" definitions. We found that for magnitudes $G \leqslant 16$, we had a FPR for the metallcity regime [Fe/H] < -2 of just 1 in 4, while this is 1 in 2 for the metallicity regime [Fe/H] < -3 for stars of magnitude $G \leqslant 14$.

However, given the exterior effective temperature and surface gravity constraints that are available we considered how the FPR varies for known intrinsic stellar parameters. We found that when considering stars with magnitude $G \leqslant 16$, for the metallicity regime [Fe/H] < -2, the FPR was less than 40$\%$ when avoiding 3500 K $< T_{\rm eff} < 4000$ K and  $T_{\rm eff} \sim 5500 K$. We additionally found that with a sufficient effective temperature and surface gravity prior, any star can be placed more specifically into regions of the Kiel diagram which allow for a clearer diagnosis of the FPR given that low-metallicity stars are frequently confined to specific isochrones. Our findings when considering stars with magnitudes $G > 16$ show very large FPRs however, we can identify regions of the Kiel diagram where the FPR is lower, $\sim 60 \%$, and hence metal-poor detections of moderate confidence are possible.

We additionally found that Gaia DR3 has the potential to drastically out-perform other low-metallicity stellar spectroscopic surveys regarding the success-rate of metal poor detections. When we adopt a similar magnitude limit to literature (magnitudes dimmer than $G = 14$) we found a success rate of metal-poor star identifications of 86\%, 67\% and 44\% for [Fe/H] $\leqslant$ -2.0, -2.5 and -3.0 respectively.

Our results suggest that Gaia DR3 has the potential to accurately identify extremely-metal-poor stars ([Fe/H] < -3), however, those attempting to identify them must be conscious of the potential for false positive detections and adjust their strategy for identification appropriately. We advise observing bright ($G \leqslant 16$) stars and preferably treating detections within the effective temperature range 3500 K $< T_{\rm eff} < 4000$ K and those within 50 K of $T_{\rm eff} \sim 5550 K$ with caution, due to the locally high FPR.

We finally conclude that uncertainties on Carbon abundance should allow for the identification of a CEMP candidate population for follow-up high-resolution spectroscopic observations. However, due to large uncertainties on $\alpha$-abundance, diagnosing $\alpha$-abundance with Gaia BP/RP will be very challenging.

\section*{Acknowledgements}

We thank the anonymous referee for providing useful comments which improved
the quality of this paper.
CECW acknowledges support from the Science and Technology Facilities Council (STFC) for a Ph.D. studentship. 
DA acknowledges support from the ERC Starting Grant NEFERTITI H2020/808240.
JLS acknowledges the support of the Royal Society (URF\textbackslash R1\textbackslash191555).
For the purpose of open access, the author has applied a Creative Commons Attribution (CC BY) licence to any Author Accepted Manuscript version arising
from this submission.
\section*{Data Availability}

The data underlying this article may be shared on reasonable request
to the corresponding author.

\bibliographystyle{mnras}
\bibliography{Ref} 

\bsp
\label{lastpage}
\end{document}